\documentclass{mem}
\usepackage{natbib}\usepackage{txfonts}\usepackage{balance}
\usepackage{graphicx}
\idline{84}{1}
\usepackage{txfonts} 

\begin{document}

\def\sna{SN\,1987A}
\def\chandra{{\em Chandra}}
\def\xmm{{\em XMM-Newton}}
\newcommand{\kms}{~{km\,s$^{-1}$}}
\newcommand{\tttt}[1]{{$\times 10^{#1}$}}
\newcommand{\hii}{H{\footnotesize \,II}}
\newcommand{\vb}{{v-b}}

\title{
\sna\ at High Resolution
}


\author{
Daniel \,Dewey  
          }

 
\institute{
MIT Kavli Institute for Astrophysics and Space Research;
70 Vassar St., Room NE80-6085;
Cambridge, MA ~02139 ~USA.~
\email{dd@space.mit.edu}
}

\authorrunning{Dewey }

\titlerunning{SN\,1987A}

\abstract{
Handed the baton from {\em ROSAT}, early observations of \sna\ with the
\chandra\ HETG and the \xmm\ RGS showed broad lines with a FWHM of
$\sim$ 10$^4$\kms: the SN blast wave was continuing to shock the \hii\
region around \sna. Since then, its picturesque equatorial ring (ER) has been
shocked, giving rise to a growing, dominant narrow-lined component. 
Even so, current
HETG and RGS observations show that a broad component is still present
and contributes $\sim$ 20\% of the 0.5--2~keV flux. \sna's X-ray behavior
can be modeled with a minimum of free parameters as the sum of two
simple 1D hydrodynamic simulations: i) an on-going interaction with
\hii\ region material producing the broad emission lines and most of
the 3--10~keV flux, and ii) an interaction with the dense, clumpy ER
material that dominates the 0.5--2~keV flux. Toward the future, we
predict a continued growth of the broad component but a drop in the
0.5--2~keV flux, {\em once} no new dense ER material is being shocked.
When? Time, and new data, will tell.
}

\maketitle{}

\section{X-ray Observations \& Modeling}

\begin{figure*}[t!]
\includegraphics[clip=true,angle=270,scale=0.50]{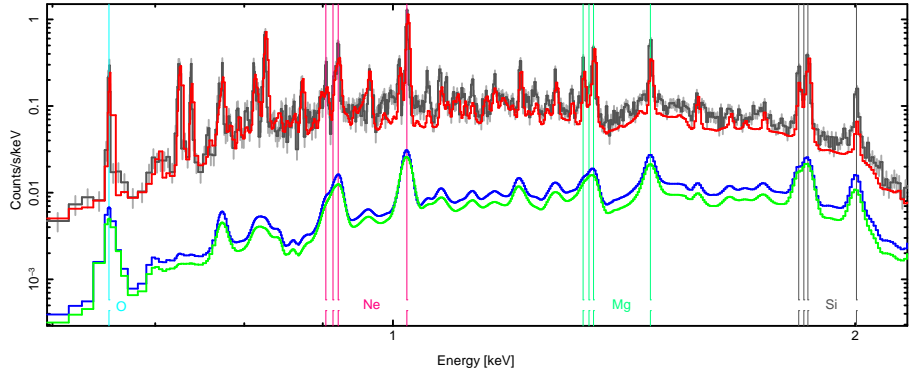}
\caption{\footnotesize
\sna's X-ray spectrum and the spectra of its hydro-model components.
The HETG/MEG data from 2011 (black with error bars) is shown along
with the modeled spectra from individual emission components;
the color-coding of components is the same
as used in Fig.~\ref{fig:side}.
The spectrum from the dense clumpy equatorial ring (ER, red)
accounts for most
of the emission in this range.
The spectra from the out-of-plane \hii\ region interactions,
shocked CSM (blue) and reverse-shocked ejecta (green),
contribute lower-flux but with very broad lines.
In future, the shocked-ejecta emission is predicted to
overtake the shocked-CSM so that 
at late times the ejecta and its abundances will dominate the
very-broad \hii\ component.
\label{fig:hydrospect}}
\end{figure*}

X-rays from \sna\ were detected by ROSAT
and attributed to the shock reaching an \hii\ region interior
to the equatorial ring (ER); in agreement,
an early \chandra\ HETG observation 
showed very broad lines \citep{Michael02}.
While monitored by \chandra\ \citep{Park06} and
\xmm\ \citep{Haberl06} the 0.5--2~keV flux dramatically increased
as the shock reached parts of the visible ER \citep{Sugerman02}.
Observations with the \chandra\ LETG \citep{Zhekov05}
were followed by further
deep grating observations as more of the ER was shocked.
Deep HETG observations \citep{Dewey12} show
narrow and broad components, Figure~\ref{fig:hydrospect};
a growing broad flux is also seen in optical observations \citep{Fransson13}.
Continued X-ray observations are being made by \xmm\ \citep{Maggi12} and
\chandra\ \citep{Helder12SN1987A}.

\begin{figure}[hb]
\begin{center}
\includegraphics[clip=true,scale=0.36]{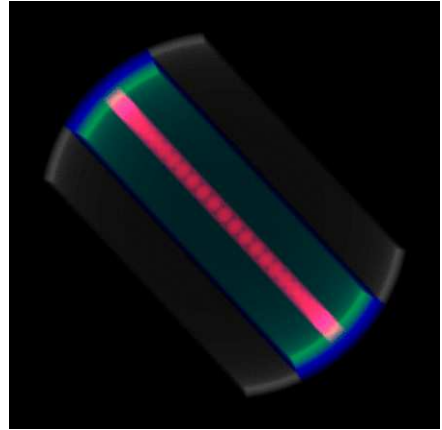}
\end{center}
\caption{
\footnotesize
Hydrodynamics-based geometric emission model of \sna\ at the meeting
epoch,
2012 October.  In this ``side'' view north is up
and the Earth is to the left.
Four emission components are color-coded: the \hii-shocked-CSM material (blue),
the reverse-shocked ejecta (green), the emission from the shocked,
clumpy ER (red), and higher-lattitude radio emission (gray).
\label{fig:side}}
\end{figure}

\sna\ can be coarsely modeled using two 1D hydrodynamic
simulations \citep{Dewey12}: one for the ER and one for
the \hii\ region.
Calculating the X-ray spectra, as in
\citet{Dwarkadas10}, gives good agreement
and explains the broad component, Figure~\ref{fig:hydrospect}.
A current schematic, Figure~\ref{fig:side}, shows that the forward
shock in the out-of-plane \hii\ region
has now passed most of the visible ER.

Monitoring \sna\
over the coming centuries
will indeed be a worthwhile
legacy project for Earthling astrophysics.

\begin{acknowledgements}
NASA provided support through SAO contract SV3-73016 to
MIT for support of the \chandra\ X-ray Center (CXC) and Science Instruments;
the CXC is operated by SAO for and
on behalf of NASA under contract NAS8-03060.
\end{acknowledgements}

\bibliographystyle{aa}


\end{document}